\documentclass[10pt,conference,onecolumn,draftcls]{IEEEtran}
\normalsize
\usepackage{amsmath,amsfonts,amssymb,amsbsy,url,verbatim}
\usepackage{comment}
\usepackage{graphicx}
\usepackage{url}
\usepackage{balance}
\usepackage{times}
\usepackage{dsfont}
\usepackage{subfigure}
\usepackage{cite}
\usepackage{bm}
\usepackage{epstopdf}
\usepackage{algorithm,algorithmic, color}
\usepackage{amsthm}
\usepackage{bm}
\usepackage{setspace}
\newtheorem{corollary}{Corollary}

\newtheorem{proposition}{Proposition}

\renewcommand{\vec}[1]{\bm{#1}}
\newcommand{\mat}[1]{\mathbf{#1}}
\newcommand{\hermitian}{^{H\scriptscriptstyle}}
\newcommand{\transpose}{^{T\scriptscriptstyle}}
\newcommand{\radixM}{^{\frac{1}{2}\scriptscriptstyle}}
\newcommand{\radixH}{^{\frac{H}{2}\scriptscriptstyle}}

\newcommand{\bi}{\begin{itemize}}
\newcommand{\ei}{\end{itemize}}
\newcommand{\ben}{\begin{enumerate}}
\newcommand{\een}{\end{enumerate}}
\newcommand{\bc}{\begin{cases}}
\newcommand{\ec}{\end{cases}}
\newcommand{\bd}{\begin{description}}
\newcommand{\ed}{\end{description}}

\newcommand{\be}{\begin{equation}}
\newcommand{\ee}{\end{equation}}
\newcommand{\bea}{\begin{eqnarray}}
\newcommand{\eea}{\end{eqnarray}}

\newcommand{\rank}{{\mathrm{rank}}}
\newcommand{\Trace}{{\mathrm{Tr}}}

\IEEEoverridecommandlockouts

\begin{document}

\title{SNOPS: Short Non-Orthogonal Pilot Sequences \\ for Downlink Channel State Estimation in FDD Massive MIMO}

\author{Beatrice~Tomasi, Alexis~Decurninge, Maxime~Guillaud\\
Mathematical and Algorithmic Sciences Laboratory, Huawei Technologies Co. Ltd.\\
France Research Center, 20 quai du Point du Jour, 92100 Boulogne Billancourt, France\\
email: \texttt{\{beatrice.tomasi,alexis.decurninge,maxime.guillaud\}@huawei.com}}

\maketitle

\begin{abstract}
Channel state information (CSI) acquisition is a significant bottleneck in the design of Massive MIMO wireless systems, due to the length of the training sequences required to distinguish the antennas (in the downlink) and the users (for the uplink where a given spectral resource can be shared by a large number of users).
In this article, we focus on the downlink CSI estimation case. Considering the presence of spatial correlation at the base transceiver station (BTS) side, and assuming that the per-user channel statistics are known, we seek to exploit this correlation to  minimize the length of the pilot sequences. We introduce a scheme relying on non-orthogonal pilot sequences and feedback from the user terminal (UT), which enables the BTS to estimate all downlink channels. Thanks to the relaxed orthogonality assumption on the pilots, the length of the obtained pilot sequences can be strictly lower than the number of antennas at the BTS, while the CSI estimation error is kept arbitrarily small.
We introduce two algorithms to dynamically design the required pilot sequences, analyze and validate the performance of the proposed CSI estimation method through numerical simulations using a realistic scenario based on the one-ring channel model.
\end{abstract}


\section{Introduction}

Channel state information (CSI) acquisition represents an important problem in the multi-user Massive MIMO (Multiple-Input Multiple-Output) scenario \cite{Larsson_MassiveMIMO_IEEE_CommMag2014}. Accurate downlink CSI is required in order to obtain the large multiplexing gain expected from massive MIMO systems and achieve the rates shown e.g. in~\cite{JSAC2013HoydisDebbah}. 
It is well known that, in the presence of i.i.d. channels, it is necessary to make the length of the pilot sequences at least as large as the total number of transmit antennas, in order to avoid the effect known as pilot contamination \cite{Ngo_Marzetta_Larsson_contamination_ICASSP2011}. Depending on the coherence time of the channel, the transmission of long training sequences instead of data-bearing symbols can represent a significant loss in spectral efficiency.
This issue is exacerbated in frequency-division duplex (FDD) systems, where downlink CSI can not be simply obtained by reciprocity \cite{isspa2005} at the BTS; in that scenario, the large number of BTS antennas makes the use of orthogonal pilot sequences especially unwelcome.

In the context of Massive MIMO however, the channels exhibit a large degree of correlation \cite{Hoydis_MMIMO_measurements_ISWCS2012}.
In fact, a denser antenna array improves the spatial resolution, and makes the received signal more spatially correlated, to the point of resulting in a rank deficient spatial correlation matrix \cite{Yin_Gesbert_etal_coordinated_estimation_JSAC2013}. This correlation can potentially help reduce the required training overhead.
The problem of pilot design for MIMO correlated channels has been studied previously, for example in~\cite{TSP2004KotechaSayeed}, \cite{TSP2010BjornsonOttersten}, \cite{Love2014} and \cite{TSP2014ShariatiBengtsson}. However, in these works the pilot optimization is done for a point-to-point MIMO system.
The single-user assumption was lifted in \cite{Adhikary_JSDM_IT2013}, where it is proposed to schedule uplink CSI acquisition across the users such that the terminals can be separated in space; pilot sequences orthogonality between users is therefore not required, yielding shorter training sequences. However, it is not clear how close to perfect separation in space a practical system can operate, considering realistic propagation conditions, and a finite number of users to choose from at the scheduling stage. 
An alternative approach was introduced in \cite{Gao_etal_common_sparsity_CSI_FDDMIMO_SP15}, where channel sparsity in the time, frequency and angular domains is assumed, and short pilots are obtained through compressed sensing techniques.

Another aspect of the problem, related to mitigating pilot contamination in the context of multi-cell Massive MIMO, has been considered in \cite{Yin_Gesbert_etal_coordinated_estimation_JSAC2013}; in this work, the same (orthogonal) pilot sequences are reused across the cells, and the problem of assigning each user to one of the pilot sequences is considered.\\

In \cite{pilot_length_minimization_Asilomar2015}, the design of non-orthogonal pilots based on per-user spatial covariance information has been addressed for \emph{uplink} multi-user Massive MIMO CSI estimation; the object of the present article is to introduce a downlink counterpart to this approach.
In the sequel, we seek to optimize the design of the training sequences transmitted by the BTS during downlink CSI estimation. The salient features of the proposed approach are:
\begin{itemize}
\item The BTS designs a set of non-orthogonal pilots based on statistical CSI, and transmits them.
\item Each UT feeds back (typically after quantization) the corresponding received sequence to the BTS.
\item The BTS reconstructs the CSI based on UT feedback and prior covariance information.
\end{itemize}
As will be seen, this enables to dynamically control the accuracy of CSI estimation, as well as to use short, non-orthogonal pilot sequences (SNOPS), suitably adapted to the channel statistics.

This article is organized as follows: we first introduce the system model and the pilot-based downlink CSI estimation model in Section~\ref{section_sysmodel}. In Section~\ref{section_optimization}, we introduce two algorithms to design pilot sequences with the objective of minimizing their length or their total energy, while simultaneously ensuring that the channel estimation error variance is upper-bounded by a per-user constant. Finally, in Section~\ref{section_simulations}, we show through simulations using a realistic scenario based on the one-ring channel model that the proposed technique can yield pilot sequences of length significantly smaller than the number of antennas at the BTS.

\begin{figure}
\centering
		\includegraphics[width=0.7\textwidth]{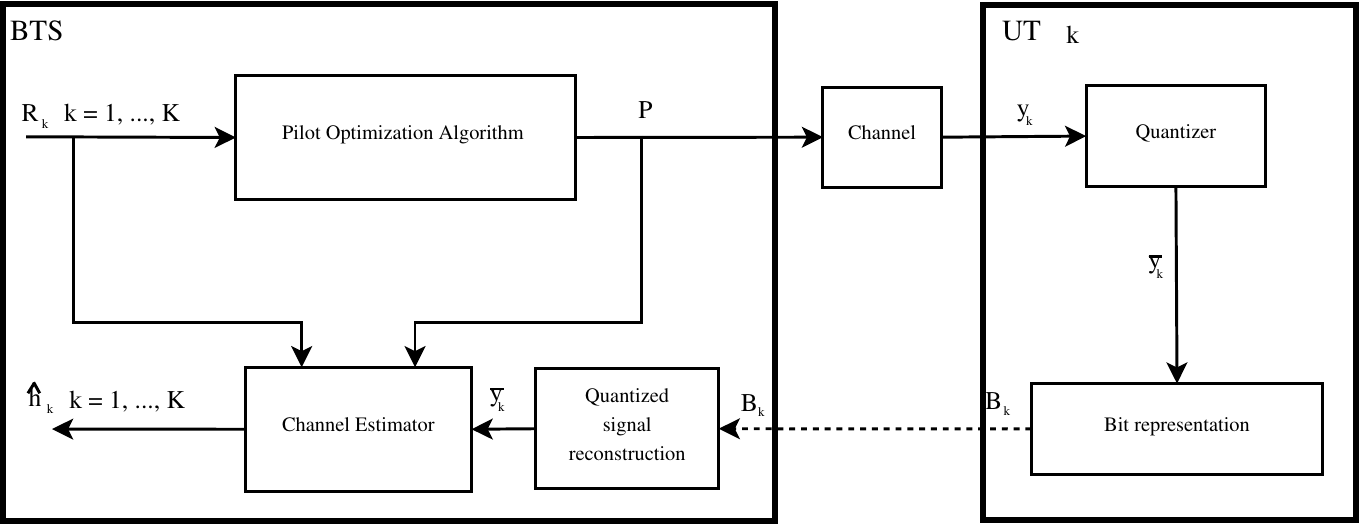}\caption{Functional representation of the proposed CSI acquisition approach. Only one among the $K$ UTs is shown.}\label{fig:pic_scheme}
	\end{figure}

\section{System Model}
\label{section_sysmodel}
\vspace{-1mm}
\subsection{Channel model}

We consider a communication system with $M$ antenna elements at the BTS and $K$ single-antenna UTs, with $M>K$. In order to incorporate spatial correlation in the model, it is assumed that the column vector of the flat-fading downlink channel coefficients between the $M$ antennas at the BTS and the $k$-th single-antenna UT, $\mat{h}_{k} \in \mathbb{C}^{M}$ can be expressed as the product between a spatial correlation matrix $\mat{R}_{k}\radixM  \in \mathbb{C}^{M\times r_k}$ and a vector $\vec{\eta}_{k} \in \mathbb{C}^{r_k}$ of complex zero-mean Gaussian i.i.d. random variables with covariance $\mat{I}_{r_k}$ that represents the fast fading process, i.e.,
\be \label{ch_model}
\mat{h}_{k} = \mat{R}_{k}\radixM \vec{\eta}_{k}.
\ee
In this model, $r_k$ is the rank of the spatial covariance matrix, $\mat{R}_{k}$,  and is lower or equal to the number of antenna elements in the BTS, $M$. We will decompose each per-user covariance matrix as $\mat{R}_{k}\radixM = \mat{U}_k\mat{\Lambda}_k\radixM$, where $\mat{\Lambda}_k$ and $\mat{U}_k$ are respectively the diagonal matrix containing the $r_k$ non-zero eigenvalues of $\mat{R}_k$, also incorporating a path loss coefficient, and the matrix of the associated eigenvectors. In this study, $\mat{R}_{k}$ is assumed to be constant (stationary channel process) and known at the BTS, i.e., statistical CSI is available\footnote{The issue of estimating the per-user BTS-side spatial covariance is outside of the scope of this paper. One approach is to apply classical covariance estimation and tracking methods, based on the estimated instantaneous CSI. Another method, based on a learned dictionary of uplink/downlink covariance matrix pairs, suitable for FDD systems, was introduced in \cite{FDD_covariance_interpolation_Globecom2015}.}.

\subsection{Pilot-Based CSI Estimation Approach}

Since we are only concerned with downlink CSI estimation, we focus on the transmission of pilot sequences by the BTS. Let us assume that pilot sequences of overall length $T$ are simultaneously broadcasted by all $M$ antennas; we let matrix $\mat{P} \in \mathbb{C}^{T \times M}$ denote the complete set of pilot sequences used in the system. The signal received at UT $k$ over the corresponding $T$ time instants, denoted by vector $\mat{y}_{k}\in \mathbb{C}^{T}$, can be expressed as:
\be 
\mat{y}_{k} = \mat{P} \mat{h}_{k}+\mat{w}_k
\ee
where $\mat{w}_k$ represents the per-user additive white Gaussian noise at UT $k$, with covariance $\sigma_{n,k}^2 \mat{I}_T$. 
Each UT will then quantize the received signal and transmit it, e.g., in the form of a binary representation $\mathrm{B}_k$, back to the BTS. We assume that the feedback link is perfect, and therefore the BTS can perfectly recover the quantized version of $\mat{y}_{k}$, for all $k = 1, \dots, K$, denoted as
\be \label{ybar_k}
\bar{\mat{y}}_{k} = \mat{y}_{k} + \mat{z}_k = \mat{P} \mat{h}_{k} + \mat{w}_k + \mat{z}_k
\ee
where $\mat{z}_k$ represents the per-user quantization error, which is assumed Gaussian with covariance $\sigma_{Q,k}^2 \mat{I}_T$ and independent from the other variables in the problem. 
The BTS can then compute the channel estimate $\hat{\mat{h}}_k$ based on the combined knowledge of $\bar{\mat{y}}_{k}$,  $\mat{R}_{k}$ and $\mat{P}$, using a simple linear MMSE approach (more details are provided in Section~\ref{section_errorcovariance} below).\\

A pictorial representation of the above steps is shown in Fig.~\ref{fig:pic_scheme}.  Some remarks about this approach are in order:
\begin{itemize}
\item The situation of interest in this paper is $T < M$; in this case, $\mat{P}$ does not admit a left inverse, and it is not possible to recover exactly $\mat{h}_k$ from $\mat{y}_k$ only, even in the noise-free case. Therefore, according to the proposed model, the UT does not directly try to estimate the channel.
\item On the other hand, the BTS is in a position to estimate $\mat{h}_k$ thanks to the statistical CSI assumed available there. In fact, in a Massive MIMO scenario the downlink CSI is used for multi-user precoding, thus making its availability more important at the BTS.
\item Due to the above, the channel statistics (in the form of spatial covariance matrices) are required at the BTS only.
\item Having short pilot sequences is doubly advantageous, since it improves the spectral efficiency of the downlink pilot transmission, and reduces the dimension of the uplink feedback variable $\bar{\mat{y}}_{k}$.
\item Although mathematically similar (but not identical), the uplink \cite{pilot_length_minimization_Asilomar2015} and downlink (presented here) CSI estimation methods differ significantly in terms of system design: the former requires the feed-forward coded transmission of the \emph{pilot sequences} $\mat{p}_{k}$ from the BTS to each UT; while the latter involves the coded feed-back of the sequence $\mat{y}_{k}$ as received by UT $k$ to the BTS.
\end{itemize}



\subsection{LMMSE Estimation Error Analysis}
\label{section_errorcovariance}

We now introduce the considered Linear Minimum Mean Square Error (LMMSE) channel estimator and derive the estimation error covariance matrix of each user, as a function of the choice of the common pilot sequences. For a given choice of the pilot matrix $\mat{P}$, the LMMSE estimator of the fast fading coefficients between user $k$ and the BTS array, $\hat{\vec{\eta}}_k$, can be written as
\be
\hat{\vec{\eta}}_k = \mat{C}_{\vec{\eta}_k\bar{\mat{y}}_k}\mat{C}_{\bar{\mat{y}}_k}^{-1}\bar{\mat{y}}_k
\ee
where  $\mat{C}_{\vec{\eta}_k\bar{\mat{y}}_k} = \mat{R}_k\radixH \mat{P}\hermitian$, and $\mat{C}_{\bar{\mat{y}}_k} = \mat{P} \mat{R}_k \mat{P}\hermitian + \sigma_k^2 \mat{I}_{T}$, where $\sigma_k^2=\sigma_{n,k}^2+\sigma_{Q,k}^2$ represents the variance of the combination of the thermal and quantization noise terms.
Considering \eqref{ch_model}, the CSI estimate is obtained for each user as
\be
\hat{\mat{h}}_k = \mat{R}_k\radixM \hat{\vec{\eta}}_k.
\ee

The covariance matrix of the estimation error for $\vec{\eta}_k$ is given by~\cite{libro_linearestimation}
\bea
\mat{C}_{\bf e,\vec{\eta}_k} (\mat{P})\!\!\!\! &=& \!\!\!\!\mathbb{E}[(\hat{\vec{\eta}}_k-\vec{\eta}_k)(\hat{\vec{\eta}}_k-\vec{\eta}_k)\hermitian]\\
\!\!\!\!&=&\!\!\!\! \mat{I}_{r_k}-\mat{R}_k\radixH \mat{P}\hermitian(\mat{P}\mat{R}_k\mat{P}\hermitian+\sigma_k^2\mat{I}_{T})^{-1} \mat{P}\mat{R}_k\radixM \label{eq:beforeKailath}\\
\!\!\!\!&=& \!\!\!\!\Big(\mat{I}_{r_k}+\frac{\mat{R}_k\radixH \mat{P}\hermitian \mat{P}\mat{R}_k\radixM}{\sigma_k^2}\Big)^{-1} \label{eq:afterKailath}
\eea
where from eq.~\eqref{eq:beforeKailath} to eq.~\eqref{eq:afterKailath} we used the Woodbury identity~\cite{matrix_cookbook}. The covariance matrix of the estimation error on $\vec{h}_k$ is therefore
\bea
\mat{C}_{{\bf e},k} (\mat{P})&=&\mathbb{E}[(\hat{\mat{h}}_k-\mat{h}_k)(\hat{\mat{h}}_k-\mat{h}_k)\hermitian]\\
&=& \mat{U}_k\mat{\Lambda}_k\radixM \mat{C}_{\bf e,\vec{\eta}_k} (\mat{P})\mat{\Lambda}_k\radixM \mat{U}_k\hermitian \\
&=&  \mat{U}_k\Big(\mat{\Lambda}_{k}^{-1}+\frac{\mat{U}_k\hermitian\mat{P}\hermitian \mat{P}\mat{U}_k}{\sigma_k^2}\Big)^{-1} \mat{U}_k\hermitian.  \label{eq_Ce_expanded}
\eea \\

\section{Minimum length pilot sequences under estimation error constraints}
\label{section_optimization}

In order to control the accuracy of the channel estimation process for user $k$, we assume that we wish to uniformly bound the estimation error on all dimensions of $\vec{h}_k$ by a given constant $\epsilon_k > 0$. This can be done by requiring that all the eigenvalues of $\mat{C}_{{\bf e},k}$ are lower than or equal to a per-user constant $\epsilon_k$, which we denote\footnote{For two positive semidefinite matrices $\mat{A}$ and $\mat{B}$, $\mat{A} \preceq \mat{B}$ is a shorthand notation for the condition that $\mat{B}-\mat{A}$ is positive semidefinite.} as $\mat{C}_{{\bf e},k} \preceq \epsilon_k\mat{I}_{M}$.
In this section, we address the design of short pilot sequences under such estimation error constraints.

\subsection{Minimum pilot length without energy constraint}
We consider the problem of minimizing the length $T$ of the pilots subject to the per-user maximum estimation error constraints outlined above, first assuming unbounded transmission energy: 
\begin{eqnarray}\label{optim_unboundedpower}
	& \underset{ \mat{P}\in \mathbb{C}^{T \times M}} \min & T  \label{optim_unbounded_P}  \\
	& \mathrm{s.t. } &  \mat{C}_{{\bf e},k} (\mat{P}) \preceq \epsilon_k \mat{I}_{M} \quad \forall k = 1, \dots, K.  \nonumber
\end{eqnarray}

We prove the following result in Appendix \ref{appendix_proof1}:
\begin{proposition}
\label{prop1}
If for each user $k$ the $r_k$ nonzero eigenvalues of $\mat{R}_k$ are greater than $\epsilon_k$, i.e., $\vec{\Lambda}_k \succeq \epsilon_k \vec{I}$, then the minimum $T$ from \eqref{optim_unbounded_P}
is $\underset{ k = 1, \dots, K}{\max} r_k$.
\end{proposition}

In other terms, in the regime of low $\epsilon_k$ (high accuracy), the minimum pilot length needed for downlink CSI acquisition is limited by the user whose covariance has the highest rank. 

\subsection{Energy-Constrained Pilot Length Minimization}

We now revisit the optimization problem \eqref{optim_unbounded_P} by introducing a maximum pilot energy constraint, i.e. we seek the minimum $T$ for which there exists a $T\times M$ matrix $\mat{P}$ that satisfies $\mat{C}_{{\bf e},k}~\preceq~\epsilon_k\mat{I}$ and that satisfies a per-antenna maximum pilot energy constraint $E_{\mathrm{max}}$:
\begin{eqnarray}\label{eq:rank1}
	& \underset{ \mat{P}\in \mathbb{C}^{T \times M}} \min & T\\
	& \mathrm{s.t. } &  \mat{C}_{{\bf e},k} \preceq \epsilon_k \mat{I}_{M} \quad \forall k = 1, \dots, K \nonumber\\
	& &  X_{mm} \leq E_{\mathrm{max}} \quad \forall m = 1, \dots, M \nonumber
\end{eqnarray}
where $\mat{X}= \mat{P}\hermitian\mat{P} \in \mathbb{C}^{M\times M}$ and $X_{mm}$ denotes the $m$-th diagonal element of $\mat{X}$. Note that minimizing $T$ is equivalent to minimizing $\rank(\mat{X})$, therefore \eqref{eq:rank1} becomes:
	\bea \label{eq_sdp1}
	&\underset{\mat{X} \succeq {\bf 0}} \min & \rank(\mat{X})\\
	&\mathrm{s.t. } &  \mat{C}_{{\bf e},k} \preceq \epsilon_k  \mat{I}_{M} \quad \forall k = 1, \dots, K\nonumber\\
	& &  X_{mm} \leq E_{\mathrm{max}} \quad \forall m = 1, \dots, M. \nonumber
	\eea

As a consequence of Proposition~\ref{prop1}, we get the following Corollary.
\begin{corollary}
\label{corollary1}
Let $\lambda_{1,k},\dots,\lambda_{r_k,k}$ denote the $r_k$ nonzero eigenvalues of $\mat{R}_k$. The optimum $T$ from \eqref{eq:rank1} is lower bounded by
\[
\underset{ k = 1, \dots, K}{\max} \sum_{i=1}^{r_k} \mathds{1}\{\lambda_{i,k}\geq \epsilon_k\}.
\]
\end{corollary}
\begin{IEEEproof}
The constraints of \eqref{eq:rank1} are more restrictive than the constraints of \eqref{optim_unbounded_P}, therefore the minimum $T$ from \eqref{eq:rank1} is greater than the optimum $T$ from \eqref{optim_unbounded_P}. Using a similar argument, the minimum $T$ from \eqref{optim_unbounded_P} is greater than the optimum $T$ from \eqref{optim_unbounded_P} where we projected each error covariance matrix on the eigenspace composed by eigenvectors whose eigenvalues are greater than $\epsilon_k$. The lower bound is then a consequence of Proposition~\ref{prop1}.
\end{IEEEproof}

The problem \eqref{eq:rank1} can efficiently be solved approximately through heuristics, e.g., following the method outlined in \cite{Fazel04rankminimization} as follows. Considering a regularized smooth surrogate of the rank function, namely $\log\det(\mat{X}+\delta \mat{I})$ for some small $\delta$, \eqref{eq_sdp1} becomes:
	\bea \label{eq_concave_objective}
	&\underset{\mat{X} \succeq {\bf 0}} \min & \log\det(\mat{X}+\delta \mat{I})\\
	&\mathrm{s.t. } &\mat{C}_{{\bf e},k} \preceq \epsilon_k \mat{I}_{M} \quad \forall k = 1, \dots, K \nonumber\\
	& &  X_{mm} \leq E_{\mathrm{max}} \quad \forall m = 1, \dots, M. \nonumber
	\eea

The objective function in \eqref{eq_concave_objective} is concave, however it is smooth on the positive definite cone; a possible way to approximately solve this problem is to iteratively minimize a locally linearized version of the objective function, i.e. solve
\begin{eqnarray}  \label{eq:rank_minimization2}  
				\mat{X}_{t+1} = \arg &  \underset{\mat{X} \succeq {\bf 0}} {\min } & \mathrm{Tr}\left((\mat{X}_t+\delta \mat{I})^{-1}\mat{X}\right)\\
				& \mathrm{s.t. } & \mat{C}_{{\bf e},k} \preceq \epsilon_k \mat{I}_{M}  \quad \forall k = 1, \dots, K\nonumber\\
	& &  X_{mm} \leq E_{\mathrm{max}} \quad \forall m = 1, \dots, M \nonumber
\end{eqnarray}
until convergence to some $\mat{X}^*$. We suggest to initialize the algorithm by choosing $\mat{X}_0$ as the rank-1, all-ones matrix $\mathbf{1}_{M\times M}$.

Let us now focus on the constraints on $\mat{C}_{{\bf e},k}, \ k = 1, \dots, K$. Using \eqref{eq_Ce_expanded} and the fact that $\mat{C}_{\bf e,k}$ is positive semidefinite, we obtain
\begin{eqnarray} \label{eq_constraint_maxerror}
\mat{C}_{{\bf e},k} &\!\!\! \preceq\!\!\!& \epsilon_k \mat{I}_{M} \ \Leftrightarrow \ \mat{U}_k\hermitian \mat{X}\mat{U}_k\succeq \left(\epsilon_k^{-1}\mat{I}_{r_k}- \mat{\Lambda}_k^{-1}\right) \sigma_k^2\\
&& \forall k = 1, \dots, K.\nonumber
\end{eqnarray}
Note that since these constraints are convex, \eqref{eq:rank_minimization2} is a convex optimization problem that can be solved numerically.

Due to the various approximations involved in transforming \eqref{eq_sdp1} into \eqref{eq:rank_minimization2}, $\mat{X}^*$ might not be strictly rank-deficient, but it can have some very small eigenvalues instead. It is therefore necessary to apply some thresholding on these eigenvalues to recover a strictly rank-deficient solution. Let us denote by $\vec{e}_m$ the eigenvector associated to the $m$-th eigenvalue $v_m$ of $\mat{X}^*$, $m=1\ldots M$, with $v_1\geq \ldots \geq v_M\geq 0$. We then let 
\begin{eqnarray} \label{eq_eigthresholding}
T&=& \max_{i=1\ldots M} i\\
&& \mathrm{s.t.} \quad v_i \geq \epsilon_s \nonumber
\end{eqnarray}
for a suitably chosen (small) $\epsilon_s$, and obtain the matrix of optimized training sequences as $\mat{P} = \mat{V}\radixM [\vec{e}_1, \dots, \vec{e}_{T}]^T$, where $\mat{V}$ is the diagonal matrix having $v_1 \ldots v_T$ as diagonal coefficients.\\

The proposed algorithm is shown in Algorithm~\ref{algo_DL_length_minimization}.

	\begin{algorithm} 
		\caption{Minimum length downlink pilot sequence computation with estimation error and per-antenna energy constraints.}
		\label{algo_DL_length_minimization}
		\begin{algorithmic}
			\STATE{Initialize $\mat{X}_0 \leftarrow \mathbf{1}_{M\times M}$.}
			\REPEAT
			\STATE{$\begin{array}{rcl} 
			\mat{X}_{t+1} &\leftarrow& \arg   \underset{\mat{X} \succeq {\bf 0}} {\min }  \mathrm{Tr}\left((\mat{X}_t+\delta \mat{I})^{-1}\mat{X}\right)\\
				&\mathrm{s.t. }&  \mat{U}_k\hermitian \mat{X}\mat{U}_k \succeq \left(\epsilon_k^{-1} \mat{I}_{r_k}- \mat{\Lambda}_k^{-1}\right) \sigma_k^2 \quad \forall k\\ 
	& &  X_{mm} \leq E_{\mathrm{max}} \quad \forall m = 1, \dots, M \nonumber
				\end{array}$ }
			\UNTIL{convergence to $\mat{X}^*$.}
			\STATE{Compute $T$ according to \eqref{eq_eigthresholding}.}
			\STATE{\textbf{Output:} $\mat{P} = \mat{V}\radixM[\vec{e}_1, \dots, \vec{e}_{T}]^T$.}
		\end{algorithmic}
	\end{algorithm}

\subsection{Pilot Energy Minimization}

We now propose an algorithm where we search the pilots minimizing the total energy dedicated by the BTS to pilot sequences, i.e. $\mathrm{Tr}(\mat{X})$, under the same CSI accuracy constraints as before. Note that the trace criterion is also a proxy criterion for the rank (see \cite{Fazel04rankminimization}), incorporating both the energy minimization and the pilot length minimization objectives. This leads to solve the convex optimization problem presented in Algorithm~\ref{algo_DL_rankM_energy_minimization}.

\begin{algorithm} 
	\caption{Pilot energy minimization with estimation error and per-antenna energy constraints.}
	\label{algo_DL_rankM_energy_minimization}
	\begin{algorithmic}

    \STATE{Compute \bea
     \mat{X}^* \leftarrow	& \arg \underset{ \mat{X} \succeq {\bf 0}} \min& \mathrm{Tr}(\mat{X}) \label{eq:optimization2}  \\
	    & \mathrm{s.t. } &  \mat{C}_{{\bf e},k} \preceq \epsilon_k \mat{I}_M \quad \forall k = 1, \dots, K\nonumber\\
	& &  X_{mm} \leq E_{\mathrm{max}} \quad \forall m = 1, \dots, M \nonumber
	\eea }
    \STATE{Compute $T$ according to \eqref{eq_eigthresholding}.}
    \STATE{\textbf{Output:} $\mat{P} = \mat{V}\radixM[\vec{e}_1, \dots, \vec{e}_{T}]^T$.}
    	\end{algorithmic}
\end{algorithm}

\section{Numerical results}
\label{section_simulations}
	
The algorithms presented in this paper have been evaluated through numerical simulations, according to the scenario outlined below: the BTS is equipped with a uniform circular array (UCA) of diameter $2$~m, consisting of $M = 32$ antennas, and serves $K = 8$ UTs randomly distributed around the BTS at a distance between $250$ and $750$~m. For each UT, it is assumed that $200$ scatterers are distributed randomly on a disc of radius $50$~m centered on each terminal are causing fast fading (one-ring channel model \cite{OneRingModel}). The covariance matrices are generated by a ray-tracing procedure, with a central frequency of $3.5$~GHz; according to this model, the support of the angle of arrivals associated to a given UT is limited, which yields covariance matrices with few large eigenvalues. We have applied a threshold to these eigenvalues to obtain the ranks $r_k$ that ensure that at least 99\% of the energy of the full-rank matrix is captured by the rank-deficient model. Distance-dependent path-loss is also incorporated, following the 3D-UMi LoS (dense urban micro-cell, line-of-sight) scenario from \cite{3GPP_TR36873_v1200}.
The noise variance (incorporating thermal noise and quantization noise, as detailed in \eqref{ybar_k}) is chosen identically across all users as $\sigma_k^2 = -110$~dBm.

Algorithms~\ref{algo_DL_length_minimization} and \ref{algo_DL_rankM_energy_minimization} have been evaluated numerically. For each realization of the covariance matrices, we applied the proposed algorithm to compute $\mat{P}$. Since the effect of the estimation error $\hat{\vec{h}}_k-\vec{h}_k$ must be interpreted relatively to the channel variance, we introduce a global accuracy parameter $\epsilon^*$, and set the per-user constraints as a function of the path-loss, according to
\be
\epsilon_k = \epsilon^* \Trace\{\mat{R}_k\},
\ee 
where $\Trace$ denotes the trace operator; note that $\Trace\{\mat{R}_k\}=\mathbb{E}\left[||\vec{h}_k ||^2\right]$. 
The solutions to \eqref{eq:rank_minimization2} and \eqref{eq:optimization2} were obtained via the numerical solver CVX \cite{cvx}, and parameter $\delta$ was set to $10^{-2}$. Moreover, for thresholding we consider $\epsilon_s = 10^{-5} \max{(v_1,\ldots, v_M)}$.

	\begin{figure}
	\centering
		\includegraphics[width=0.7\textwidth]{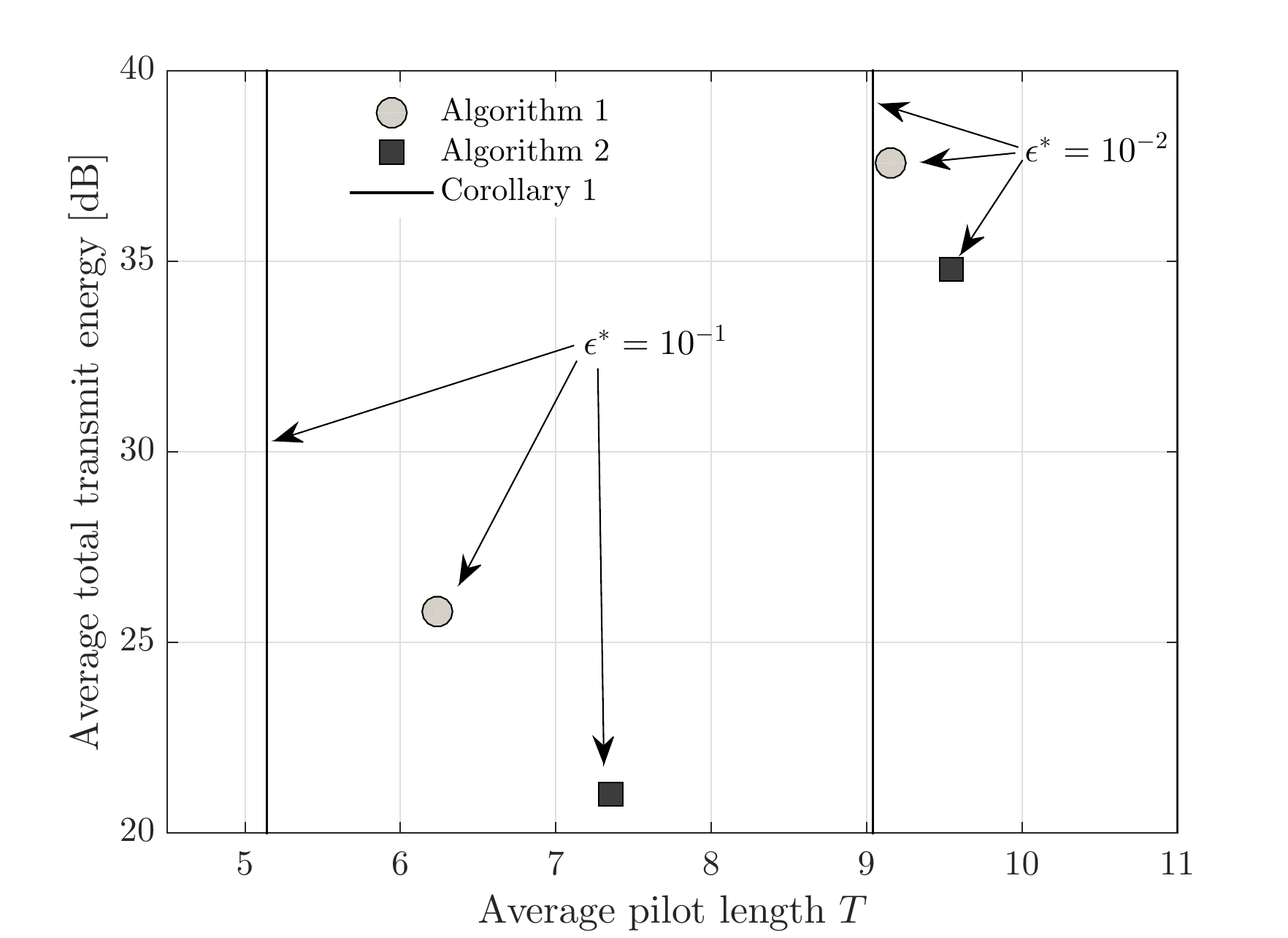}\caption{Average total transmit energy and average pilot length obtained by Algorithms~\ref{algo_DL_length_minimization} and~\ref{algo_DL_rankM_energy_minimization} in a system with $M=32$ antennas at the BTS.}\label{fig:energy_rank}
	\end{figure}
	
Fig.~\ref{fig:energy_rank} depicts the total transmit energy contained in the pilot sequences transmitted by the BTS versus the corresponding length $T$, as obtained by Algorithm~\ref{algo_DL_length_minimization} in light gray and Algorithm~\ref{algo_DL_rankM_energy_minimization} in dark gray. The maximum energy constraint per antenna, $E_{\mathrm{max}}$, is set to $41$~dB. The vertical lines represent the average lower bound on the minimum rank of the pilot sequence matrix derived in Corollary~\ref{corollary1}. All metrics are averaged over $100$ realizations of the random user locations. It can be seen that Algorithm~\ref{algo_DL_length_minimization} performs closer to the optimal (the loss of optimality being attributable to a combination of the per-antenna energy constraint and the influence of the user eigenvalues below the precision threshold $\epsilon_k$ that are not taken into account in the lower bound), while Algorithm~\ref{algo_DL_rankM_energy_minimization} has better (lower) energy figures, at the cost of slightly longer sequences. It is noticeable that the obtained pilot sequences are significantly shorter (average length lower than 10 for $\epsilon^* = 10^{-2}$) than what would be necessary in the absence of statistical CSI, i.e. $T=M=32$. 

\begin{figure}
\centering
		\includegraphics[width=0.7\textwidth]{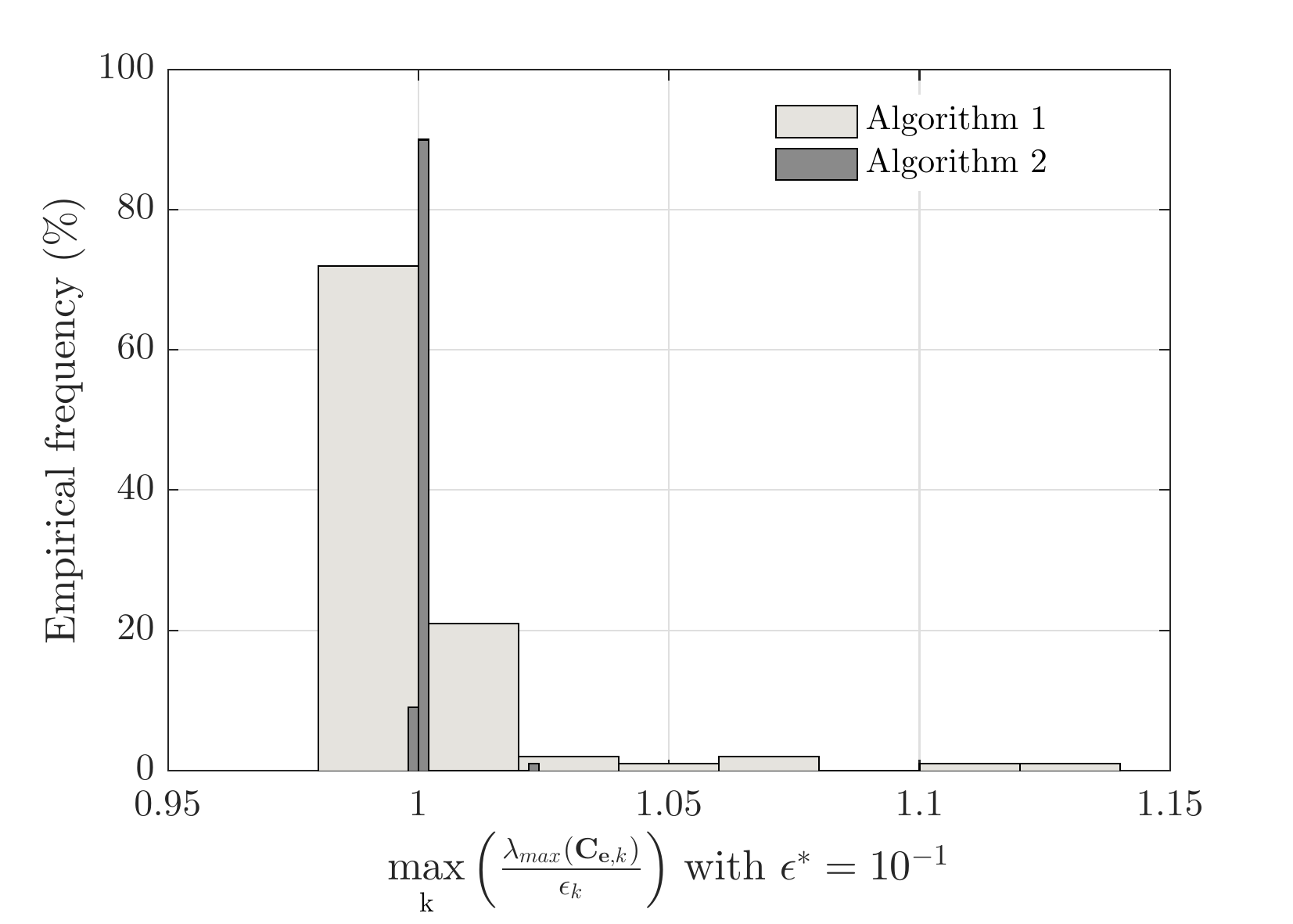}\caption{ Empirical frequency of the maximum value of the ratio between the maximum eigenvalue of the error covariance matrix and the accuracy constraint for Algorithms~\ref{algo_DL_length_minimization}  and~\ref{algo_DL_rankM_energy_minimization}, when $\epsilon^* = 10^{-1}$.}\label{fig:histo1}
	\end{figure}

\begin{figure}
	\centering
		\includegraphics[width=0.7\textwidth]{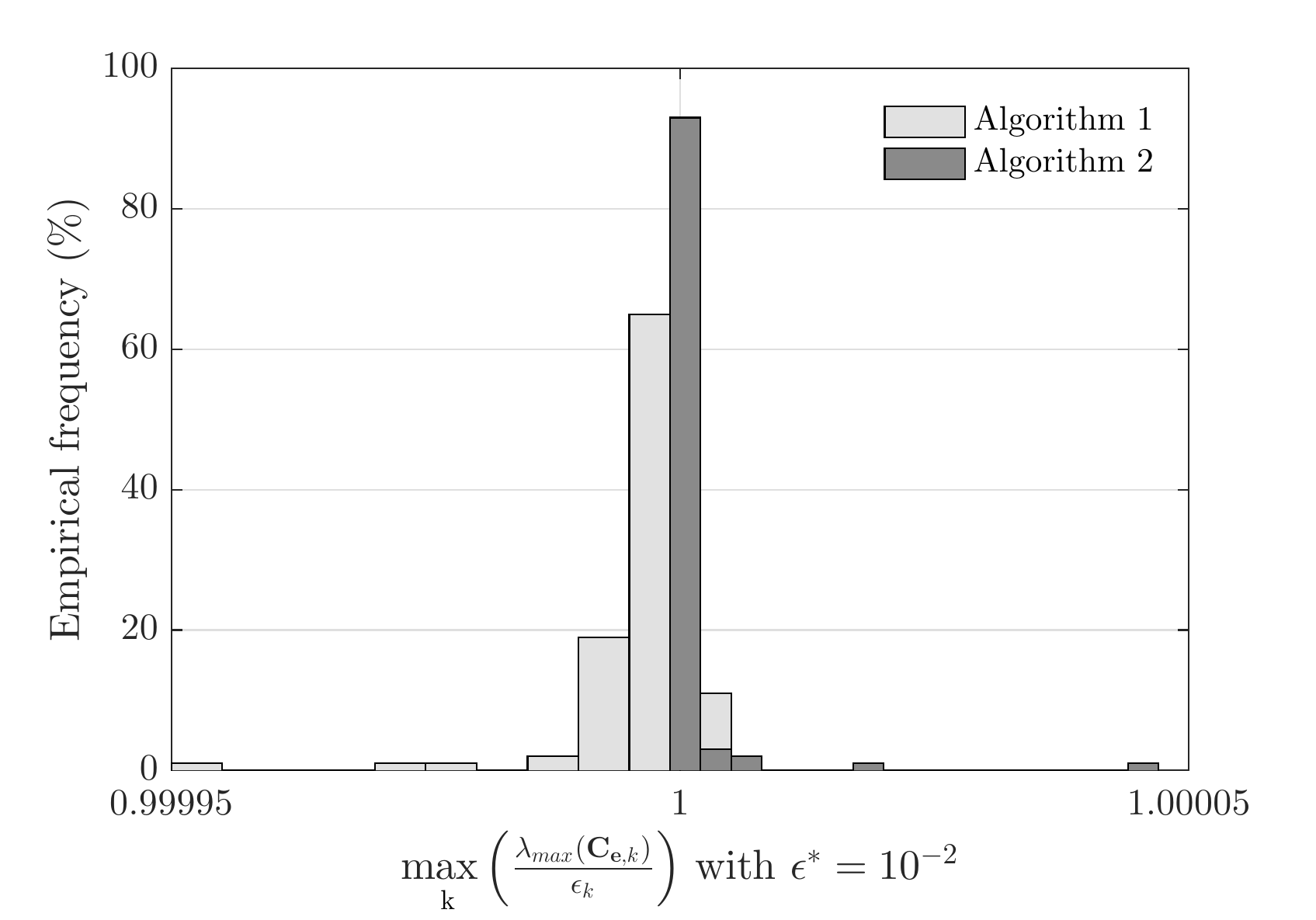}\caption{ Empirical frequency of the maximum value of the ratio between the maximum eigenvalue of the error covariance matrix and the accuracy constraint for Algorithms~\ref{algo_DL_length_minimization}  and~\ref{algo_DL_rankM_energy_minimization}, when $\epsilon^* = 10^{-2}$.}\label{fig:histo2}
	\end{figure}

Because of the thresholding operation detailed in \eqref{eq_eigthresholding}, for both Algorithm~\ref{algo_DL_length_minimization} and Algorithm~\ref{algo_DL_rankM_energy_minimization}, $\mat{X}= \mat{P}\hermitian\mat{P}$ is only approximately equal to $\mat{X}^*$. Therefore, the obtained pilot sequences can slightly violate the maximum error variance constraints $\mat{C}_{{\bf e},k} \preceq \epsilon_k \mat{I}_M$.
In order to verify that this has no detrimental effect on the global performance, let us consider the quantity $\max_k \lambda_{\mathrm{max}}(\mat{C}_{{\bf e},k})/\epsilon_k$, which should always be below 1 in the absence of the thresholding operation.
Figs.~\ref{fig:histo1} and~\ref{fig:histo2} show the empirical distributions of these quantities for parameter $\epsilon^* = 10^{-1}$ and $\epsilon^* = 10^{-2}$, respectively. These results show that the departure from the ideal value of 1 is extremely small, and therefore the effect of the thresholding operation on the final CSI estimation error is negligible.


\section{Conclusions}
In this paper we addressed the problem of downlink CSI acquisition in a MIMO FDD single-cell cellular network. We proposed two algorithms for pilot sequence design that exploit the higher spatial correlation of the channel coefficients between the BTS antennas and the UTs. Pilot sequences shorter than the current state of the art solutions can be employed to estimate the channel coefficients with a desired accuracy. We evaluated analytical bounds for the rank of pilot matrices and we evaluated numerically the proposed algorithms.

\begin{appendices}

\section{Proof of Proposition~\ref{prop1}}
\label{appendix_proof1}
Considering eq.~\eqref{eq_Ce_expanded}, let us denote $
\mat{Z}_k =\mat{P}\mat{U}_k$.
Proposition~\ref{prop1} follows as consequence of the following propositions proved in this Appendix.
\begin{proposition}
Assume that $\mat{\Lambda}_k \succeq \epsilon_k \mat{I}$ for all users $k = 1,\dots, K$, then solving the minimization problem \eqref{optim_unboundedpower}
is equivalent to solving 
\begin{eqnarray} 
	& \underset{ \mat{P}\in \mathbb{C}^{T \times M}} \min & T \label{minlength_individual_rank}\\
	& \mathrm{s.t. } &  \rank(\mat{Z}_k) = r_k \quad \forall k = 1, \dots, K. \nonumber
\end{eqnarray} 
\end{proposition}

\begin{IEEEproof}
Let us first recall that $\mat{C}_{\bf e,k}\preceq \epsilon_k \mat{I}$ is equivalent to
$\mat{\Lambda}_k^{-1}-\frac{\mat{I}}{\epsilon_k} + \frac{\mat{Z}_k\hermitian\mat{Z}_k}{\sigma_k^2}\succeq 0$.
The matrix $\mat{D}_k=\frac{\mat{I}}{\epsilon_k} -\mat{\Lambda}_k^{-1}$ is diagonal of size $r_k\times r_k$ with positive coefficients.\\
Let us first suppose that there exists $\mat{P}$ such that $\mat{C}_{\bf e,k}\preceq \epsilon_k \mat{I}$. Then for any $\vec{x}\in\mathbb{C}^{r_k}$,
$\|\mat{Z}_k\vec{x}\|^2-\sigma_k^2\|\mat{D}_k\vec{x}\|^2\geq 0$.
Then, for any $\vec{x}\neq 0$, it holds $\|\mat{Z}_k\vec{x}\|\geq \sigma_k^2\|\mat{D}_k\vec{x}\|^2>0$, i.e. $\mat{Z}_k$ is full rank, equal to $r_k$.\\
Conversely, let us suppose that there exists $\mat{P}$ such that $\mat{Z}_k$ is full rank. Then, let us consider $\mat{P}_2=\alpha \mat{P}$ with some power level $\alpha>0$. We indicate as $\mat{Z}_{2,k} = \mat{P}_2 \mat{U}_k$.
It holds that for any $\vec{x}\in\mathbb{C}^{r_k}$, $ \|\mat{Z}_{2,k}\vec{x}\|^2-\sigma_k^2\|\mat{D}_k\vec{x}\|^2\geq \alpha^2\lambda_{min}(\mat{Z}_k\hermitian\mat{Z}_k)-\sigma_k^2\lambda_{max}(\mat{D}_k^2)$.
Since $\mat{Z}_k$ is full rank, it holds $\lambda_{min}(\mat{Z}_k\hermitian\mat{Z}_k)>0$ for all $1\leq k\leq K$. Then by choosing $\alpha$ such that
$\alpha^2\geq \max_{1\leq k\leq K} \frac{\sigma_k^2\lambda_{max}(\mat{D}_k^2)} {\lambda_{min}(\mat{Z}_k\hermitian\mat{Z}_k)}$,
it holds  that $\frac{\mat{Z}_{2,k}\hermitian\mat{Z}_{2,k}}{\sigma_k^2}- \mat{D}_k\succeq 0 \quad \forall k$.
\end{IEEEproof}

\begin{proposition}
The minimum $T$ attained in \eqref{minlength_individual_rank}
is equal to $\max_{1\leq k\leq K} r_k$.
\end{proposition}
\begin{IEEEproof}
It is straightforward to prove that if $\rank(\mat{Z}_k) = r_k$ then 
$r_k = \rank(\mat{Z}_k) = \rank(\mat{P} \mat{U}_k) \leq  \min(T,r_k)$.
Therefore, $T \geq r_k$, for all $k = 1,\dots, K$, thus, $T\geq \max(r_1, \dots, r_K)$.\\

In order to show the proposition, we will now show the existence of a pilot sequence of length $\max(r_1,...r_K)$ such that $\text{rank}(\mat{Z}_k) = r_k$ for all $1\leq k \leq K$.\\
If we denote $\text{span}(\mat{U}_k)^{\perp}$ the subspace of $\mathbb{C}^M$ orthogonal to the subspace spanned by $\mat{U}_k$, it holds
\bea
r_k\geq \text{rank}(\mat{Z}_k) &=&  \text{rank}(\mat{P}\mat{U}_k)\geq   \text{rank}(\mat{P}_{r_k}\mat{U}_k) \nonumber\\ &\geq&  r_k - \dim(\text{span}(\mat{P}_{r_k}\transpose)\cap \text{span}(\mat{U}_k)^{\perp}) \nonumber
\eea
where $\mat{P}_{\ell}$ denotes the concatenation of the $\ell$ first rows of $\mat{P}$. It is then sufficient to prove the existence of $\mat{P}$ such that for each $k$,  $\text{rank}(\mat{P}_{r_k}\mat{U}_k) = r_k$, i.e. $\text{span	}(\mat{P}_{r_k}\transpose)\cap \text{span}(\mat{U}_k)^{\perp}= \emptyset$. With such a $\mat{P}$, it will hold that $\text{rank}(\mat{Z}_k) =r_k$ for all $k$.\\
We then recursively build the pilot sequence (i.e. the rows of matrix $\mat{P}$). For any step $\ell\geq 1$, let $\text{span}(\mat{P}_{\ell-1})^{\perp}$ denote the subspace of $\mathbb{C}^M$ orthogonal to $\vec{p}_1,...,\vec{p}_{\ell-1}$. We take the convention $\text{span}(\mat{P}_{0})^{\perp}=\mathbb{C}^M$, so that the construction is valid for $\ell=1$. Let us suppose that we built $\ell-1$ pilots. Then, we choose $\vec{p}_{\ell}\in \text{span}(\mat{P}_{\ell-1})^{\perp}$ such that for all users $k$ satisfying $r_k\geq \ell$, $\text{span}(\mat{P}_{r_k})\cap \text{span}(\mat{U}_k)^{\perp}= \emptyset$. After $\ell=\max(r_1,\dots, r_K)$ steps, it holds that, for each $k$, $\text{span}(\mat{P}_{r_k}\transpose)\cap \text{span}(\mat{U}_k)^{\perp}= \emptyset$ by construction.\\
It now remains to show the existence of such $\vec{p}_{\ell}$. For each $k$ satisfying $r_k\geq \ell$, it holds $\dim(\text{span}(\mat{U}_k)^{\perp})=M-r_k<M-\ell+1=\dim(\text{span}(\mat{P}_{\ell-1})^{\perp})$.\\
 We deduce that for each k such that $r_k\geq \ell$, the subspace $\text{span}(\mat{U}_k)^{\perp}$ has an empty interior in $\text{span}(\mat{P}_{\ell-1})^{\perp}$ (see e.g. \cite{singerthorpe}), then the union on $k$ (denoted by $\mat{V}_{\ell}$) also has an empty interior. Therefore, the interior of $\text{span}(\mat{P}_{\ell-1})^{\perp}\setminus \mat{V}_{\ell}$ is not empty.
This means there exists some $\vec{p}_{\ell}\in \text{span}(\mat{P}_{\ell-1})^{\perp}$ that is not in this union which ends the proof.

\end{IEEEproof}

\end{appendices}
\balance
\bibliographystyle{IEEEtran}
\bibliography{IEEEabrv,refen2}

\begin{thebibliography}{10}
\providecommand{\url}[1]{#1}
\csname url@samestyle\endcsname
\providecommand{\newblock}{\relax}
\providecommand{\bibinfo}[2]{#2}
\providecommand{\BIBentrySTDinterwordspacing}{\spaceskip=0pt\relax}
\providecommand{\BIBentryALTinterwordstretchfactor}{4}
\providecommand{\BIBentryALTinterwordspacing}{\spaceskip=\fontdimen2\font plus
\BIBentryALTinterwordstretchfactor\fontdimen3\font minus
  \fontdimen4\font\relax}
\providecommand{\BIBforeignlanguage}[2]{{%
\expandafter\ifx\csname l@#1\endcsname\relax
\typeout{** WARNING: IEEEtran.bst: No hyphenation pattern has been}%
\typeout{** loaded for the language `#1'. Using the pattern for}%
\typeout{** the default language instead.}%
\else
\language=\csname l@#1\endcsname
\fi
#2}}
\providecommand{\BIBdecl}{\relax}
\BIBdecl

\bibitem{Larsson_MassiveMIMO_IEEE_CommMag2014}
E.~G. Larsson, F.~Tufvesson, O.~Edfors, and T.~L. Marzetta, ``Massive {MIMO}
  for next generation wireless systems,'' \emph{IEEE Commun. Mag.}, vol.~52,
  no.~2, pp. 186–--195, 2014.

\bibitem{JSAC2013HoydisDebbah}
J.~Hoydis, S.~ten Brink, and M.~Debbah, ``Massive {MIMO} in the {UL/DL} of
  cellular networks: How many antennas do we need?'' \emph{IEEE Journal on
  Selected Areas in Communications,}, vol.~31, no.~2, pp. 160--171, 2013.

\bibitem{Ngo_Marzetta_Larsson_contamination_ICASSP2011}
H.~Ngo, T.~L. Marzetta, and E.~G. Larsson, ``Analysis of the pilot
  contamination effect in very large multicell multiuser {MIMO} systems for
  physical channel models,'' in \emph{Proc. International Conference on
  Acoustics, Speech and Signal Processing (ICASSP)}, 2011, pp. 3464--3467.

\bibitem{isspa2005}
M.~Guillaud, D.~T.~M. Slock, and R.~Knopp, ``A practical method for wireless
  channel reciprocity exploitation through relative calibration,'' in
  \emph{Proc. Eighth International Symposium on Signal Processing and Its
  Applications (ISSPA '05)}, Sydney, Australia, 2005.

\bibitem{Hoydis_MMIMO_measurements_ISWCS2012}
J.~Hoydis, C.~Hoek, T.~Wild, and S.~ten Brink, ``Channel measurements for large
  antenna arrays,'' in \emph{Proc. IEEE International Symposium on Wireless
  Communication Systems (ISWCS)}, 2012.

\bibitem{Yin_Gesbert_etal_coordinated_estimation_JSAC2013}
H.~Yin, D.~Gesbert, M.~Filippou, and Y.~Liu, ``A coordinated approach to
  channel estimation in large-scale multiple-antenna systems,'' \emph{IEEE
  Journ. Sel. Areas in Comm.}, vol.~31, no.~2, pp. 264–--273, 2013.

\bibitem{TSP2004KotechaSayeed}
J.~H. Kotecha and A.~M. Sayeed, ``Transmit signal design for optimal estimation
  of correlated {MIMO} channels,'' \emph{IEEE Transactions on Signal
  Processing}, vol.~52, no.~2, pp. 546--557, Feb. 2004.

\bibitem{TSP2010BjornsonOttersten}
E.~Bj\"ornson and B.~Ottersten, ``A framework for training-based estimation in
  arbitrarily correlated {Rician MIMO} channels with {Rician} disturbance,''
  \emph{IEEE Transactions on Signal Processing}, vol.~58, no.~3, pp.
  1807--1820, 2010.

\bibitem{Love2014}
J.~Choi, D.~J. Love, and P.~Bidigare, ``Downlink training techniques for {FDD}
  massive {MIMO} systems: Open-loop and closed-loop training with memory,''
  \emph{IEEE Journal of Selected Topics in Signal Processing}, vol.~8, no.~5,
  pp. 802--814, 2014.

\bibitem{TSP2014ShariatiBengtsson}
N.~Shariati, J.~Wang, and M.~Bengtsson, ``Robust training sequence design for
  correlated {MIMO} channel estimation,'' \emph{IEEE Transactions on Signal
  Processing,}, vol.~62, no.~1, pp. 107--120, 2014.

\bibitem{Adhikary_JSDM_IT2013}
A.~Adhikary, J.~Nam, J.~Ahn, and G.~Caire, ``Joint spatial division and
  multiplexing: The large-scale array regime,'' \emph{IEEE Transactions on
  Information Theory}, vol.~59, no.~10, pp. 6441--6463, 2013.

\bibitem{Gao_etal_common_sparsity_CSI_FDDMIMO_SP15}
Z.~Gao, L.~Dai, Z.~Wang, and S.~Chen, ``Spatially common sparsity based
  adaptive channel estimation and feedback for {FDD} massive {MIMO},''
  \emph{IEEE Transactions on Signal Processing}, vol.~63, no.~23, pp.
  6169--6183, 2015.

\bibitem{pilot_length_minimization_Asilomar2015}
B.~Tomasi and M.~Guillaud, ``Pilot length optimization for spatially correlated
  multi-user {MIMO} channel estimation,'' in \emph{Proc. Asilomar Conference on
  Signals, Systems and Computers}, Pacific Grove, CA, USA, 2015.

\bibitem{FDD_covariance_interpolation_Globecom2015}
A.~Decurninge, M.~Guillaud, and D.~Slock, ``Channel covariance estimation in
  massive {MIMO} frequency division duplex systems,'' in \emph{Proc. IEEE
  Global Telecommunications Conference (GLOBECOM)}, San Diego, CA, USA, 2015.

\bibitem{libro_linearestimation}
T.~Kailath, A.~H. Sayed, and B.~Hassibi, \emph{Linear Estimation}.\hskip 1em
  plus 0.5em minus 0.4em\relax Prentice Hall, 2000.

\bibitem{matrix_cookbook}
K.~B. Petersen and M.~S. Pedersen, ``The matrix cookbook,'' Feb. 2007.

\bibitem{Fazel04rankminimization}
M.~Fazel, H.~Hindi, and S.~Boyd, ``Rank minimization and applications in system
  theory,'' in \emph{Proc. American Control Conference (ACC)}, 2004, pp.
  3273--3278.

\bibitem{OneRingModel}
D.~Shiu, G.~J. Foschini, M.~J. Gans, and J.~M. Kahn, ``Fading correlation and
  its effect on the capacity of multielement antenna systems,'' \emph{IEEE
  Transactions on Communications}, vol.~48, no.~3, pp. 502--513, 2000.

\bibitem{3GPP_TR36873_v1200}
{3GPP Technical Specification Group Radio Access Network}, ``Study on {3D}
  channel model for {LTE} (release 12),'' Tech. Rep. TR 36.873, 2015.

\bibitem{cvx}
M.~Grant and S.~Boyd, ``{CVX}: Matlab software for disciplined convex
  programming, version 2.1,'' \url{http://cvxr.com/cvx}.

\bibitem{singerthorpe}
I.~M. Singer and J.~A. Thorpe, \emph{Lecture Notes on Elementary Topology and
  Geometry}.\hskip 1em plus 0.5em minus 0.4em\relax Scott Foresman \& Co,
  Glenview, IL, 1967.

\end{thebibliography}
\end{document}